# Review of Linpack and Cloudsim on VMM

Bohar Singh, M.Tech student
SBSSTC,Ferozepur,India

Pawan Luthra, Assistant professor,
SBSSTC, Ferozepur, India

**ABSTRACT**

Virtualization is a framework of dividing the resources of a computer into multiple execution environments which offers a lot of benefits including flexibility, security, ease to configuration and reduction of cost but at the same time it also brings a certain degree of performance overhead. Furthermore, Virtual Machine Monitor (VMM) is the core component of virtual machine (VM) system and its effectiveness greatly impacts the performance of the whole system. This review paper will try to describe the basic knowledge about various virtual machine monitors such as VMware and VirtualBox. It also discussed and explores the benchmark LINPACK and CloudSim available for cloud computing. This benchmark and CloudSim can be used to measure the performance of two different virtual machine monitors in terms of processing speed, time, bandwidth, quality and response of the cloud computing network.

**Keywords:**
LINPACK, CloudSim, Virtual Machine Monitor (VMM), Virtual Machine, VMware and VirtualBox.

## 1. INTRODUCTION

Cloud computing provides better services for hosting and access to shared resources over the Internet [1, 2]. Cloud computing is best for business owners to grow from lower resources to large due to increasing demand of service. This technological trend has enabled the realization of a new computing model called cloud computing, in which user can access services from anywhere, at anywhere over the Internet in an on-demand fashion [2].

The use and deployment of clouds have many attractive benefits, such as scalability and reliability, reduced cost, elasticity, quality of service (QoS); however, development of clouds to provide more economical solutions as consumers only need to pay for what resources they required while providers can capitalize poorly utilized resources. From a provider's point of view, the maximization of the profit is major concern [3].

Many years ago, a problem invoked that how to run multiple operating systems on the same machine simultaneously? The solution to this problem was virtual machines created by VMM.

Virtual machines monitor the core part of virtual machines sits between one or more operating systems and the hardware and gives the illusion to each running OS that it controls the machine. Virtual Machine technology starts to focus on virtualization which enables to run multiple OS environments simultaneously on the same physical hardware in strong isolation from each other. Virtualization can be used to perform isolation, consolidation and migration of resources in cloud computing. VMs share the conventional hardware in a secure manner with help of VMM which control guest OS on host, while each VM is hosting its own operating system and applications [4, 5].

Indeed, the VMM serves as an operating system for operating systems, but at a much lower level, the OS must still think it is interacting with the physical hardware. Thus, transparency is a major goal of VMMs [5].

Today virtual machine has become popular due to various reasons. Server consolidation [6] is one such reason, in which virtualization enables an administrator to consolidate multiple operating systems onto fewer hardware platforms which leads to lower cost and ease administration [10]. In many setting, user run services on different machines which run different operating systems, and yet each machine is lightly utilized and isolated from another machine.

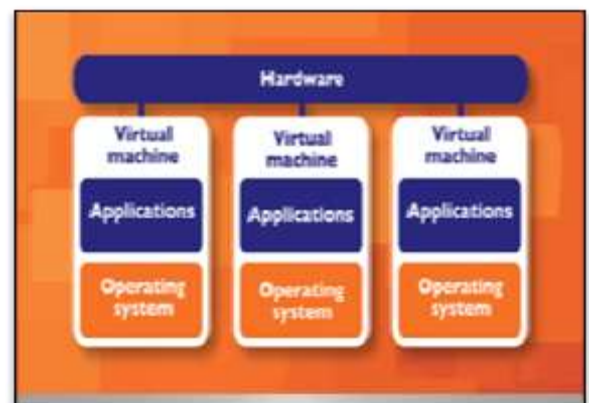

**Figure1: Overview of VM Architecture**

Virtualization has also become popular on desktops, as many users want to run one operating system but still have access to native applications on a different platform. Another reason is testing and debugging. While developers write code on one main platform, they often want to debug and test it on the many different platforms that they deploy the software to in the field. Thus, virtualization makes it easy to do so, by enabling a developer to run many different operating system types and versions on just one machine.

Section 2 discus two virtual machine monitors for virtual machines. First is VMware and second one is VirtualBox. Section 3 explores benchmark Linpack and CloudSim.  This benchmark is used to evaluate time required by machine to solve linear and least square problem whereas CloudSim is used to measure the performance of two different virtual machine monitors in terms of processing speed, time, bandwidth, network quality and response.

## 2. VIRTUAL MACHINE MONITORS (VMM)
### 2.1 VMware

VMware [7] is a virtualization and cloud computing software provider for x86 compatible computer architecture. VMware Inc. is a subsidiary of EMC Corporation, well known in the field of system virtualization and cloud computing. VMware's software allows users to create multiple virtual environments, or virtual computer systems, on a single computer or server. Essentially, one computer or server could be used to host, or manage, many virtual computer systems, sometimes as many as one hundred or more. The software virtualizes hardware components such as the video card, network adapters, CPU, memory and hard drive called hardware virtualization which is very useful for enterprisers by setting up multiple server systems on single server or physical hardware without having to purchase separate hardware for each of them. User can create virtual servers using VMware's software, leads to saving a lot of time and money [9].

VMware released its first software program, called VMware Workstation. In 2011, VMware entered the cloud computing market by releasing Cloud Foundry, an open source platform-as-a-service (PAAS) software system. This cloud computing system was designed to support applications built on Java, Ruby on Rails, Sinatra, and as well as provide support for MySQL, MongoDB, and other database platforms.

Virtual machine (VM)  is heart of virtualization, tightly isolated software container with an operating system and application inside. Because each virtual machine is completely separate and independent, many of them can run simultaneously on a single computer. A thin layer of software called a hypervisor decouples the virtual machines from the host and dynamically allocates computing resources to each virtual machine as needed.

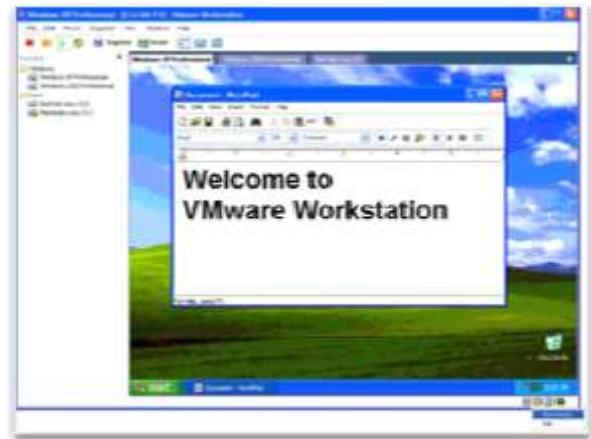

**Figure 2: A snapshot of VMWare**

A hosted x86 virtualization monitor which can run a guest operating system unmodified with some performance loss. The x86 architecture offers four levels of privilege known as Ring 0, 1, 2 and 3 to operating systems and applications to manage access to the computer hardware. While user level applications typically run in Ring 3, the operating system needs to have direct access to the memory and hardware and must execute its privileged instructions in Ring 0 [5]. Virtualizing of x86 architecture require placing a virtualization layer under the operating system to create and manage the virtual machines that deliver shared resources. Some sensitive instructions cannot effectively be virtualized as they have different semantics when they are not executed in Ring 0. The difficulty in trapping and translating these sensitive and privileged instruction requests at runtime was the challenge that originally made x86 architecture virtualization look impossible so VMware resolved this problem by developing binary translation techniques that allow the VMM to run in Ring 0 for isolation and performance, while moving the operating system to a user level ring with greater privilege than applications in Ring 3 but less privilege than the virtual machine monitor in Ring 0. It does not support Hyper Threading3 and requires a host operating system, which means an extra layer and additional overhead.

### 2.2 VirtualBox
VirtualBox [8] is a cross-platform virtualization application, installs on the existing Intel or AMD-based computers, whether they are running Windows, Mac, Linux or Solaris operating systems. Secondly, it extends the capabilities of a user's existing computer so that it can run multiple operating systems inside multiple virtual machines at the same time. A user can run Windows and Linux on Mac, run Windows Server 2008 on Linux server, run Linux on Windows PC. User can install and run many virtual machines, but the only practical limits are disk space and memory.

VirtualBox is simple and easy to use, yet also very powerful to manage VM. It can run everywhere from small embedded systems or desktop class machines. Oracle VirtualBox is an x86 cross platform open source virtualization software package developed by Oracle Corporation as part of its family of virtualization products. Virtual Box is also called hosted hypervisor, host OS is required on which VMM is installed. To a very large degree, VirtualBox is functionally identical on all of the host platforms and same file, image formats are used. The following screenshot shows VirtualBox is installed on a Mac computer, and running Windows 7 in a virtual machine as guest OS.

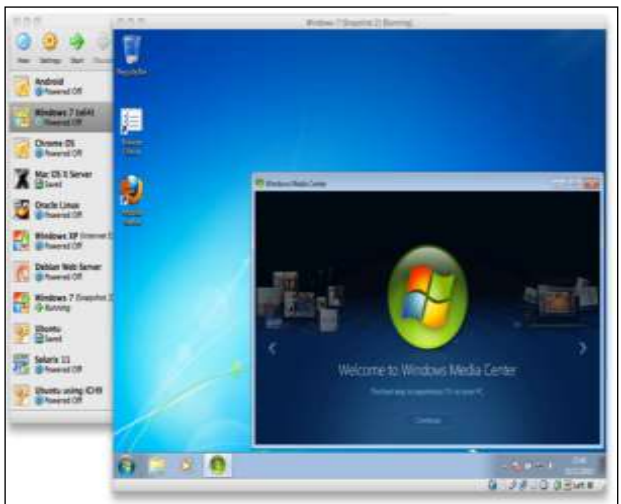

**Figure 3: A snapshot of VirtualBox**

User can run virtual machines created on one host to another host with a different host operating system. The primarily uses of VirtualBox is software virtualization to run virtual machines. This is the default behavior for any virtual machines created within the VirtualBox environment. It provides the option to enable hardware virtualization on a per virtual machine basis when running on AMD-V and Intel-VT capable CPUs. On more recent CPU designs, VirtualBox is also able to make use of nesting paging tables to improve virtual machine performance.

## 3. BENCHMARK
### 3.1 LINPACK (LINear system PACKage)
The LINPACK package is a collection of FORTRAN subroutines for solving various systems of linear equations and linear least-squares problems. The software in LINPACK is based on a de-compositional approach to numerical linear algebra. It provides a range of options to solve linear systems whose matrices are general, banded, symmetric indefinite, symmetric positive definite, triangular, and tridiagonal square, also has capability to computes the QR and singular value decompositions of rectangular matrices and applies them to least-squares problems. LINPACK uses column-oriented algorithms to increase efficiency by preserving locality of reference. The LINPACK package was based on another package which is called Level 1 Basic Linear Algebra Subroutines (BLAS) library. BLAS [13] carried out most of the floating-point work within the LINPACK algorithms. BLAS also work out with different linear equations and linear least-squares problems and makes it possible to take advantage of special computer hardware [11-14].

In the LINPACK Benchmark, a matrix of size 100 was originally used because of memory limitations with the computers. Such a matrix has 10,000 floating-point elements and could have been accommodated in most environments of that time [14]. This was done so users could estimate the time required to solve their matrix problem by extrapolation. Over the year's additional performance data was added, with increasing number of computers, the scope of the benchmark has also expanded.

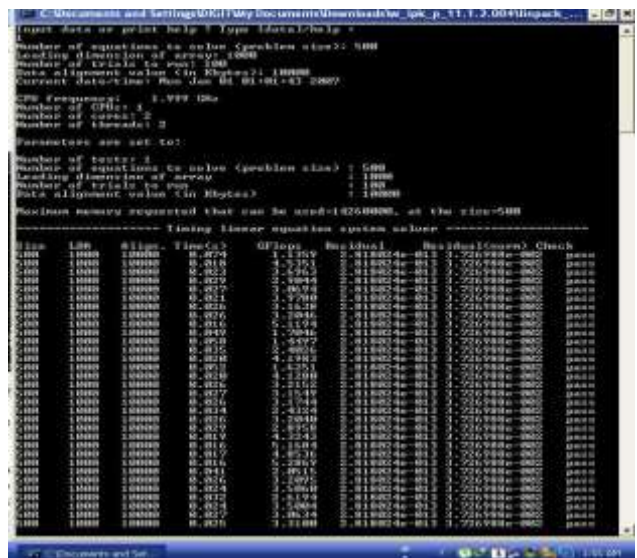

**Figure 4: A sample output from Linpack 100**

LINPACK was designed for applying to supercomputers in the early 1980s and now acts as one of the most authoritative benchmarks in high performance computers. The TOP 500 computers in the world are sorted by the LINPACK's result. To follow the development of computer architectures, LINPACK evolves into EISPACK and LAPACK.

EISPACK mainly dedicates to numerical computation of the Eigen values and eigenvectors of matrices. LINPACK measures the actual peak value of float-point computing power indicated in giga of float-point operations per second (GFLOP) [11, 12].

### 3.1.1 Linpack Variant

LINPACK 100 is very similar to the original benchmark published in 1979 and obtained by Gaussian elimination with partial pivoting with $2/3n^3 + 2n^2$ floating point operations where n is 100, the order of the dense matrix A that defines the problem. Its small size and the lack of software flexibility do not allow most modern computers to reach their performance limits. However, it can still be useful to predict the performances in a numerically intensive user written code using compiler optimization. LINPACK 1000 can provide a performance nearer to the machine's limit because to offering a bigger problem size, a matrix of order 1000, changes in the algorithm are possible.

The previous benchmarks are not suitable for testing parallel computers. For that Linpack's Highly Parallel Computing benchmark or HP Linpack benchmark was introduced. In HP Linpack the size n of the problem can be made as large as it is needed to optimize the performance results of the machine.

### 3.2 CloudSim

CloudSim is a simulation application which enables seamless modeling, simulation and experimentation of cloud computing and application services. The performance evaluation of cloud provisioning policies, services, application workload, models and resource performance models under varying system, user configurations and requirements is big challenge, to overcome this challenge, CloudSim can be used which is new, generalized and extensible simulation toolkit that enables seamless modeling, simulation and experimentation of the emerging cloud computing system, infrastructures and application environments for single and internetworked clouds. In simple words, CloudSim is a development toolkit for simulation of Cloud scenarios.

CloudSim is not a framework as it does not provide a ready to use environment for execution of a complete scenario with a specific input. Instead, users of CloudSim have to develop the Cloud scenario, to get the required output, and with defined input parameters. The CloudSim toolkit supports system and behavior modeling of cloud system components such as data centers, virtual machines and resource provisioning policies. Figure 5 shows graphic interface of CloudSim 3.0

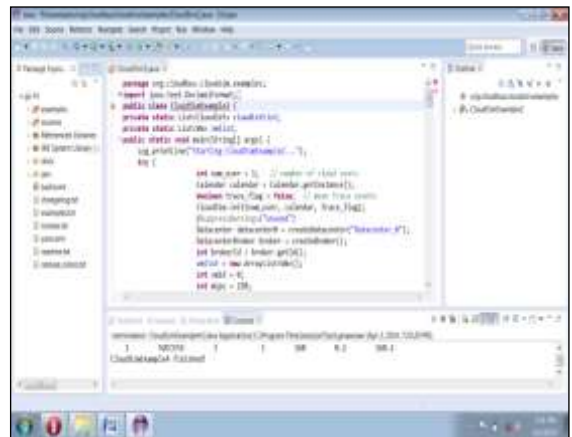

**Figure 5: A Snapshot from CloudSim 3.0**

It implements generic application provisioning techniques that can be extended with ease and limited efforts. CloudSim is an open source web application that launches preconfigured machines designed to run common open source robotic tools, robotics simulator Gazebo. SimJava is a toolkit for building working models of complex systems. It is based around a discrete event simulation kernel at the lowest level of CloudSim. It includes facilities for representing simulation objects as animated icons on screen. Figure 6 indicates output after simulation.

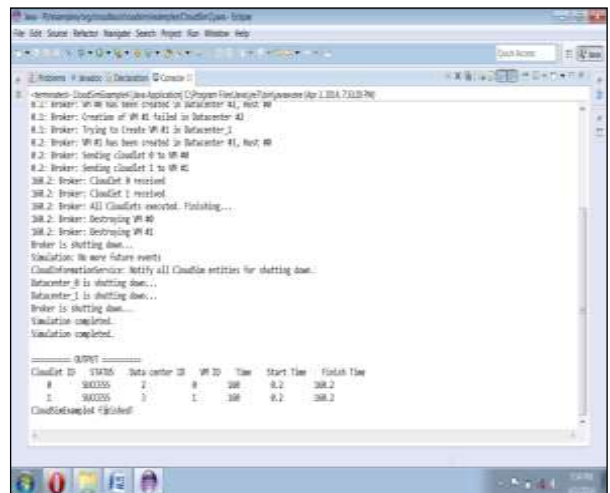

**Figure 6: A sample output from CloudSim 3.0**

## 4. CONCLUSION AND FUTURE SCOPE

Firstly, I described the basic knowledge about VMware and VirtualBox and benchmark LINPACK, CloudSim. In the future, an experiment can be conducted to measure the performance of virtualized XP on VMware and VirtualBox. It will also measure the processing speed on the basis of the number of floating point operations per second and the network response on the basis of bandwidth in TCP connection and the quality of the network in UDP connection. I will describe recent trends in high-performance computing which, I believe, will shape the near future of the LINPACK and CloudSim Benchmark suite. The results which will produce can be used for comparison.